\documentclass[aps,showpacs,twocolumn,floats,prd,superscriptaddress,nofootinbib]{revtex4} 
\usepackage{graphicx,amsmath,amssymb,amstext}
\usepackage{amssymb,amsbsy,amsfonts,amsthm,color}

\usepackage{epsfig}
\usepackage{graphicx}
\usepackage{subfigure}

\graphicspath{{Figures/}}

\begin{document}

\title{Strong lensing interferometry for compact binaries}

\author{Ue-Li Pen}
\email{pen@cita.utoronto.ca}
\affiliation{Canadian Institute for Theoretical Astrophysics, University of Toronto, M5S 3H8 Ontario, Canada, and Canadian Institute for Advanced Research CIFAR program in Gravitation and Cosmology.}

\author{I-Sheng Yang}
\email{isheng.yang@gmail.com}
\affiliation{IOP and GRAPPA, Universiteit van Amsterdam, Science Park 904, 1090 GL Amsterdam, Netherlands.}

\begin{abstract}
We propose a possibility to improve the current precision measurements on compact binaries. When the orbital axis is almost perpendicular to our line of sight, a pulsar behind its companion can form two strong-lensing images. These images cannot be resolved, but we can use multi-wavelength interferometry to accurately determine the passage through superior conjunction. This method does not depend strongly on the stability of the pulse profile, and applies equally well to both slow and fast pulsars. We discuss the possible improvement this can bring to the bound on stochastic gravitational wave background and to determine black hole spin. We also discuss the possibility of discovering a suitable binary system by the Square Kilometer Array that our method can apply to.
\end{abstract}

\maketitle

\section{Introduction}

High precision timing of compact binaries has been one of the best tools to study general relativity. The post-Keplerian orbital changes provided excellent confirmation of gravitational wave emission \cite{TayWei82,KraSta06,WeiNic10}. More recently, it was pointed out that the same data can be used to bound the stochastic gravitational wave background at an orbital frequency $\sim10^{-4}Hz$, and such bound can be lowered to the interesting range if the timing precision can be improved by a few orders of magnitude \cite{HMY}.

Current techniques to achieve high timing precision rely on a stable millisecond pulsar in the binary system. After establishing an average pulse profile, one can look for Doppler shifts due to the orbital motion. Mapping out such periodic behavior of the pulse arrival times can identify the occurrence of a particular orbital phase, for example the periastron passage. For the Hulse-Taylor system, PSR B1913+16, the current technique of millisecond pulsar timing can reach an orbital phase resolution $(\delta T_{\rm periastron}/P_{\rm orbit})\sim10^{-6}$.

In this paper, we will explore an alternative method that may provide a better precision. Our method also requires a pulsar, but unlike traditional timing techniques, it does not depend directly on the stability nor the short period of the pulses. For us, a pulsar serves as a good source of radial waves, $f\sim GHz$, $\lambda\sim10^{-1}m$. We exploit such a short wavelength in an interferometer setup, in which the two paths of light interfering with each other come from a strong lensing effect \cite{Mao92}. After decomposing the signal into multiple frequency channels, we get a 2-dimensional interference pattern \cite{WalKoo08} on which the passage through superior conjunction (when the pulsar is right behind its companion) is a special line. Recognizing such line is quite robust against the fluctuations of single pulse behavior such as microstructures \cite{Bar78,LanKra98} or jitters \cite{OslStr11}. Given a solar mass binary with a period of a few hours, the Einstein radius $R_{\rm lens}$ is about $10^6m$. Observing such interference pattern then provides a native resolution of the orbital phase that is about $(\lambda/R_{\rm lens}) \sim 10^{-7}$, which might be further improved by a thorough signal-to-noise analysis.

The lensing effect requires the binary to have a small inclination. Along our line of sight, the pulsar needs to go behind its companion within the Einstein radius. Note that the radius of a white dwarf is about $10^7m>R_{\rm lens}$, so it blocks the strong lensing signals. The pulsar's companion has to be a neutron star or a black hole. There is one known example, the double pulsars PSR J0737-3039A/B, that seems to have a small enough inclination. Unfortunately, the magnetospheres of these two pulsars are rather large, $\sim10^7m>R_{\rm lens}$, so the observation is about eclipsing instead of strong-lensing \cite{Lyu04,BreKas12}. There is not yet a clear theoretical reason why the magnetosphere has to be this large, so in the near future we might hope to discover pulsar-neutron star binaries, or simply pulsar-black hole binaries that meet our requirements \cite{SKA}. 

Before such a binary is found, our method is futuristic. The exact precision depends on the signal-to-noise analysis which cannot be predicted beforehand. In this paper, we will first go over the basic idea of this strong-lensing-interferometer. We then calculate the projected precision and discuss some practical issues to show that a pulsar binary can indeed reach it. Finally, we discuss the improvement this might bring to the upper-bound of stochastic gravitational wave provided by the binary-resonance detector \cite{HMY}, and the possibility to determine the spin of the companion neutron star or black hole.

\section{Basic Idea}

Consider a binary system with a circular orbit of radius $R_{\rm orbit}$. On the plane perpendicular to the line of sight, we can parametrize the projected orbit as 
\begin{equation}
D_{\rm proj} = R_{\rm orbit}
\sqrt{\sin^2\phi_{\rm orbit} + 
\cos^2\phi_{\rm orbit}\sin^2\theta_{\rm tilt}}~.
\end{equation}
In this toy model, the binary includes a point source of EM wave, and its partner is purely a gravitational lens. $\phi_{\rm orbit}$ is the orbital phase and we will focus on the time during which it is small, which is define to be the time when the signal source is near the superior conjunction (the farthest point behind the lens). $\theta_{\rm tilt}$ is the tilt of the orbit, and we will also focus on the cases of a small tilt such that the orbital axis is almost perpendicular to the line of sight.

\begin{figure}[tb]
\begin{center}
\includegraphics[width= 6cm]{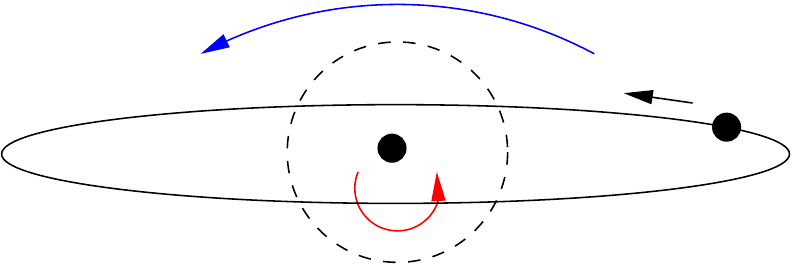}
\caption{The binary orbit in the rest frame of the lens and viewed from a small inclination angle. When the source goes behind the lens and enters the Einstein radius (the dotted circle), we can see both a far image (blue) and a close image (red).}
\label{fig-orbit}
\end{center}
\end{figure}

In this case, it is possible to have a strong lensing effect near the superior conjunction. We will use the unit that $G=1$, so the mass of the lens equals to its Schwarzschild radius. A far-away observer can define the Einstein radius as
\begin{equation}
R_{\rm lens} = \sqrt{M R_{\rm orbit}}~.
\end{equation}
When $D_{\rm proj} \lesssim R_{\rm lens}$, there will be two strongly lensed images with magnifications \cite{Mao92,PenBro13}
\begin{eqnarray}
\mu_\pm = \frac{u^2+2}{2u\sqrt{u^2+4}} \pm \frac{1}{2}~,
\end{eqnarray}
where $u = (D_{\rm proj}/R_{\rm lens})$. This can only happen when $\theta_{\rm tilt}\lesssim (R_{\rm orbit}/R_{\rm lens})$. The orbit and the lensing images can be visualized in Fig.(\ref{fig-orbit}).

In practice, the binary is so far away and these two images cannot be resolved. What we really observe is an overall magnification\footnote{This diverges at $u=0$ when the source is right behind the lens, because that produces a full Einstein ring instead of two images. This will only happen for $\theta_{\rm tilt}$ extremely close to zero and we will not address such a rare situation.}
\begin{equation}
\mu_{\rm total} = \mu_++\mu_- = \frac{u^2+2}{u\sqrt{u^2+4}}~.
\end{equation}
This leads to a feature on the signal amplitude within a range of orbital phases:
\begin{equation}
|\phi_{\rm orbit}| \lesssim \Delta\phi_{\rm amp} 
= \frac{R_{\rm lens}}{R_{\rm orbit}}~.
\end{equation}
Thus, observing this feature already implies a native resolution up to $\Delta\phi_{\rm amp}$, but that is not the main effect we would like to use here. 

\begin{figure}[tb]
\begin{center}
\includegraphics[width= 8cm]{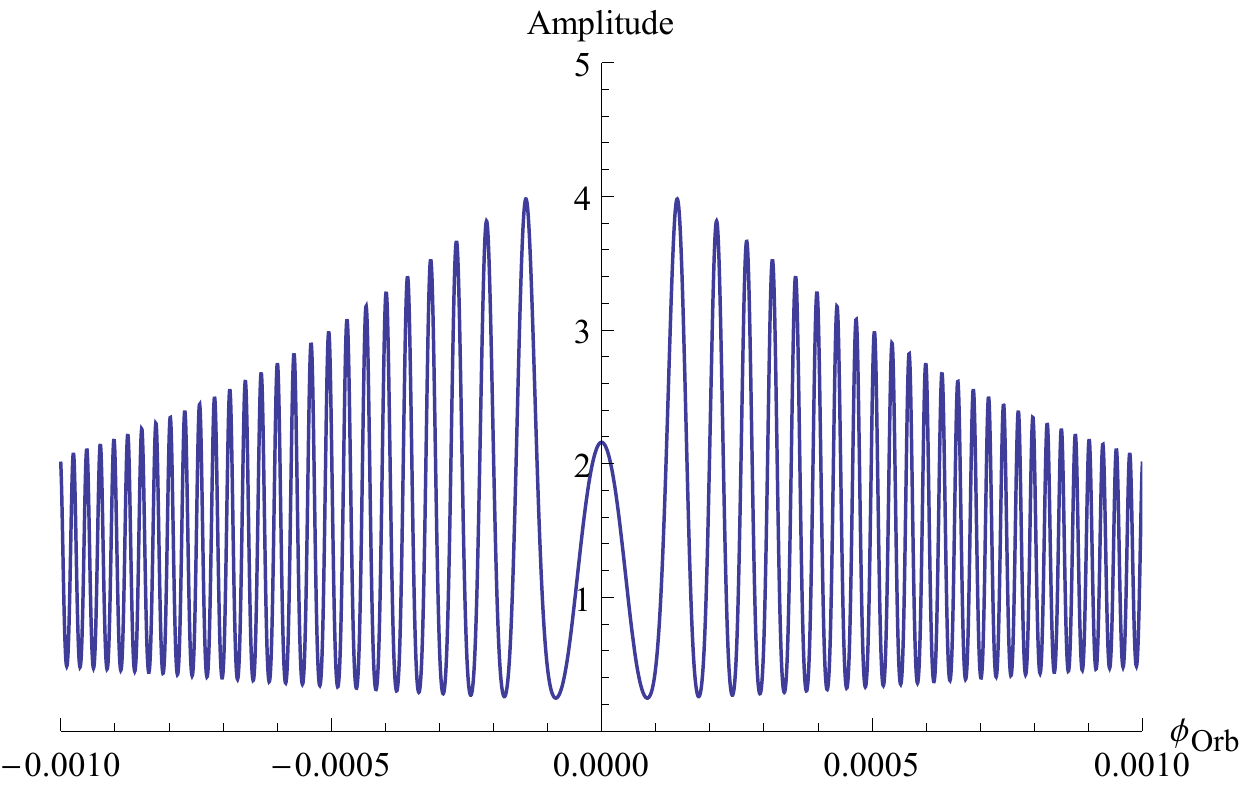}
\caption{The amplitude profile as a function of orbital phase $\phi_{\rm orbit}$. The orbital parameters are $R_{\rm orb} = 10^9m,~R_{\rm lens} = 10^6m$, and $\theta_{\rm tilt}=0.5(R_{\rm lens}/R_{\rm orb})$. We chose $\lambda=50m$, which is about $f=6MHz$, such that both the magnification envelope and the interference pattern are visible. The $GHz$ observation will have a denser interference pattern (by a factor of hundreds).}
\label{fig-int}
\end{center}
\end{figure}

When the EM wave is monochromatic, these two images will interfere with each other. The length difference between their paths is given by \cite{Mao92,PenBro13}
\begin{equation}
\Delta l = M 
\left(\frac{u\sqrt{u^2+4}}{2} + \ln\frac{\sqrt{u^2+4}+u}{\sqrt{u^2+4}-u}\right)~.
\label{eq-pathdiff}
\end{equation}
The intensity as a function $\phi_{\rm orbit}$ should really be an interference pattern.
\begin{equation}
I(\phi_{\rm orbit}) = 
\mu_+ + \mu_- + 2\sqrt{\mu_+\mu_-}
\cos 2\pi\frac{\Delta l}{\lambda}~.
\label{eq-int}
\end{equation}
We plot this pattern in Fig.(\ref{fig-int}). Qualitatively, this is a double-slit with a separation $\sim R_{\rm lens}$. Therefore, the interference pattern provides a native resolution up to
\begin{equation}
\Delta\phi_{\rm int} \sim \frac{\lambda}{R_{\rm lens}}~.
\label{eq-resint}
\end{equation}

\section{Application}

Let us put in some real numbers to estimate the expected precision of our method. For convenience, we use orbital parameters comparable to that of the Hulse-Taylor binary \cite{TayWei82}.
\begin{eqnarray}
M = 10^3 m~, \ \ \ \ 
R_{\rm orbit} = 10^9m~, \ \ \ \
R_{\rm lens} = 10^6m~.
\label{eq-orbit_parameter}
\end{eqnarray}
This leads to an orbital period of a few hours, and we will use $P_{\rm orbit}=10^4s$ for simplicity. The lensing magnification profile leads to a resolution $\Delta\phi_{\rm amp}\sim10^{-3}$. In other words, in every orbital period the signal will be magnified for 
\begin{equation}
\Delta T_{\rm amp} 
= P_{\rm orbit}\frac{\Delta\phi_{\rm amp}}{2\pi}
\sim {\rm a \ few \ seconds}~. 
\end{equation}
The interference pattern for a radio wave $f= 1GHz$ leads to $\Delta\phi_{\rm int}=3\times10^{-7}$. Near the center, the duration of each interference peak is about
\begin{equation}
T_{\rm int} = P_{\rm orbit}\frac{\Delta\phi_{\rm int}}{2\pi}
\approx 5\times10^{-4}s~.
\end{equation}
Recognizing these peaks in the data leads to the desired improvement of timing precision. The mass is a little bit of an underestimation for black holes, so the real situation might be slightly better.

Up to this point, we have pretended that the radio wave comes from a continuous, isotropic and monochromatic point source. A real pulsar is significantly different from this idealization. We will proceed to address these differences and show that our method is applicable.

\subsection{Multiple channels}

A real pulsar emission is broad-band, and different wavelengths lead to different interference patterns. The typical phase difference between two light paths during the strong lensing effect is 
\begin{equation}
\left\langle\frac{\Delta l}{\lambda}\right\rangle 
= \frac{\Delta\phi_{\rm amp}}{\Delta\phi_{\rm int}}
= \frac{M}{\lambda}\approx 3\times10^3~.
\end{equation}
This means that a narrow-band filter with $\Delta f \sim 10^{-4} f = 10^5Hz$ is required to get a clear pattern. 

\begin{figure*}[t]
 \begin{center}
  \includegraphics[width=\textwidth]{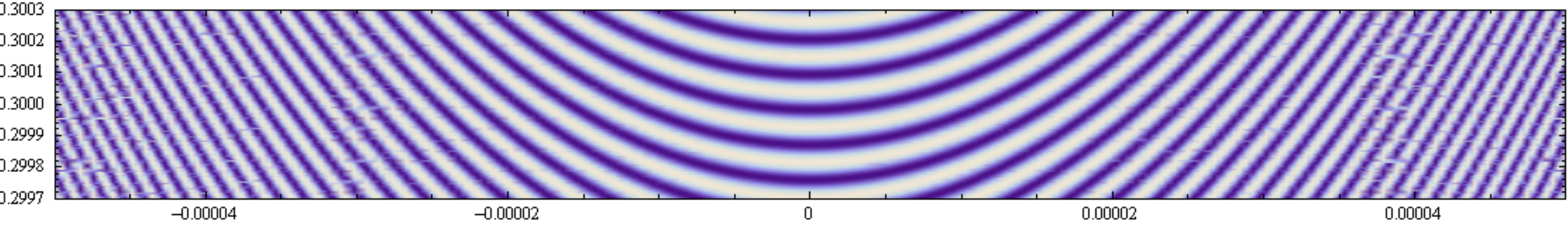}
  \includegraphics[width=\textwidth]{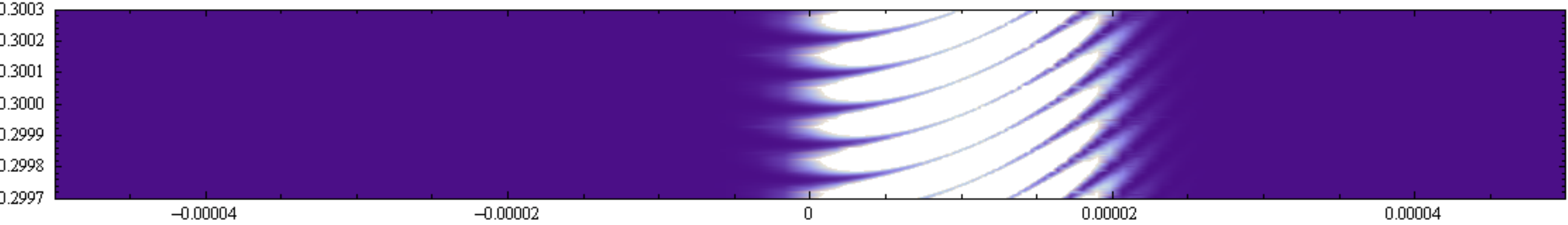}
	\includegraphics[width=\textwidth]{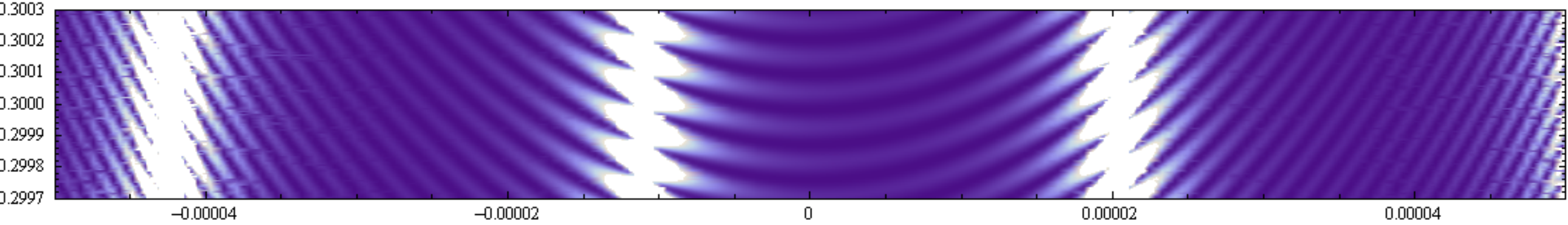}
 \end{center}
  \caption{The top panel is the standard 2-dimensional interference pattern. White regions are peaks. The vertical axis is wavelength in the unit of meter, which scans over frequency channels $(1\pm10^{-3})GHz$. The horizontal axis is the orbital phase $\phi_{\rm orb}$, and the range is $10\%$ of the full magnification envelope $\Delta\phi_{\rm amp}$. The middle panel shows how a long pulse illuminates a segment near the center. The bottom panel shows how multiple short pulses illuminate different vertical lines, along which we will get several 1-dimensional interference patterns. We assume a uniform intensity across channels.}
	\label{fig-channels}
\end{figure*}

The typical observation at $GHz$ is broad-hand, $\Delta f\sim GHz$. We can decompose it into $\sim10^4$ narrow-band channels, and their interference patterns will be correlated. This technique is similar to the secondary spectrum of timing data that reconstructs multi-path scattering events \cite{2012MNRAS.421L.132P,WalKoo08}. As shown in Fig.(\ref{fig-channels}), the interference peaks form a family of parabolas on a time-wavelength dynamic spectrum, and their common center corresponds to the superior conjunction.

Note that the narrow-band observation is essentially the original broad-band observation with a longer Fourier integration time. The $\Delta f\sim GHz$ observation can have a time-resolution up to nanoseconds, but every narrow-band channel requires a Fourier integral over $\sim10\mu s$. Fortunately, this time-resolution is enough to see the interference pattern which is about a millisecond wide. Since information is conserved, the loss of time-resolution is the only consequence of decomposing into narrow-band channels, and we do not otherwise lose $S/N$. If individual pulses have enough $S/N$ to be seen at the original broad-band observation, we will have enough narrow-band $S/N$ in multiple channels. We are simply rearranging the same information in the conjugate Fourier domain where they are more localized.


\subsection{Finite pulse duration}

Typically, the pulse duration is about $(1/10)\sim(1/100)$ of the pulse period. Since that comes from the neutron star's spinning motion, it implies that the emission has an opening angle of a few degrees \cite{Cor79}. In order to have two lensing images, the emission needs to cover the entire gravitational lens at the same time. That means $(R_{\rm lens}/R_{\rm orbit})$ has to be less than a few degrees, which is satisfied (within a factor of 10) by Hulse-Taylor-like orbits.

Although the emission is wide enough to cover the entire lens, they do not last long enough to illuminate the entire interference pattern. To see this effect we should first qualitatively put the pulsars into two categories: slow pulsars with $P_{\rm pulse}\sim$ seconds and pulse duration $10^{-1}\sim10^{-2}s~$; fast pulsars with $P_{\rm pulse}\sim$ milliseconds and pulse duration $10^{-1}\sim10^{-2}ms$. 

In the first case, we are looking for pulses which happens to be emitted when the pulsar is near the superior conjunction. Such a pulse will illuminate a large segment of the interference pattern. As shown in Fig.(\ref{fig-channels}), if such segment includes the center or is sufficiently close to the center, we can locate the center up to the projected precision $\Delta\phi_{\rm int}\sim10^{-7}$. Although slow pulsars are know to have microstructures \cite{Bar78,LanKra98}, those have no reason to behave in a similarly coherent way across multiple channels. Thus it should not be difficult to tell them apart using multi-channel techniques \cite{WalKoo08,GolGal12}. It is typical for slow pulsars to have a single-pulse $S/N\gtrsim1$, so they are ideal for our purpose. 

For fast pulsars, a pulse duration is comparable (or shorter) than the interference peak, but we will get many pulses. This means that the full pattern is illuminated in several places but for shorter durations. As shown in Fig.(\ref{fig-channels}), every single pulse illuminates a vertical line, along which we have a secondary interference pattern in the Fourier space---as we scan through different frequency channels. The separation of the peaks in this secondary pattern is largest for pulses near the superior conjunction, and it decreases as we move away. A simple interpolation can locate the maximum with a native precision similar to $\Delta\phi_{\rm int}$. Pulsar jitters \cite{OslStr11} might affect the brightness and $S/N$ for individual pulses, but the structure of the secondary interference pattern is immune to such noise. Although single-pulse $S/N$ is typically small for fast pulsars, we are using many of them together to determine one superior conjuction. Thus it is still possible to get a precise measurement.

\subsection{Emission region}

If the radiation comes from a region of finite size, then the interferometer cannot determine the source position better than its size. That is because the interference pattern coming from two different points of the source will be shifted by an amount comparable to the peak separation, thus destroying the overall pattern. In our case, the length dimension of the emission region cannot be larger than $(R_{\rm orbit}\Delta\phi_{\rm int})\sim10^2m$. Currently, there are upper-bounds of a few kilometers \cite{JohGwi12,PenMac14}\footnote{For special substructure within giant pulses, the upper bound is a few meters\cite{CrabNature}.}, but there are no lower-bounds. Thus one can still hope that this condition will be satisfied. Actually, if we see a lensing magnification without the interference pattern, it might set a lower-bound on the size of the emission region.

\subsection{Interstellar Scintillation}

In addition to multi-path propagation due to gravitational lensing, pulsars are known to undergo multi-path scattering due to structures in the interstellar medium. The latter leads to prominent scintellation, which can be undone using interstellar holography \cite{2014MNRAS.440L..36P,2008MNRAS.388.1214W}.  It is unlikely to lead to confusion with gravitational lensing which has the orbital periodicity and a very specific pattern as we demonstrated.

\section{Discussion}

The binary-resonance effect can provide an upper-bound on the stochastic gravitational wave background at the orbital frequency of a binary \cite{HMY}. Such upper-bound is roughly given by
\begin{equation}
h_c \sim 5 \left(\frac{\delta T}{P_{\rm orbit}}\right)
n_{\rm data}^{1/2} 
\left(\frac{T_{\rm tot}}{P_{\rm orbit}}\right)^{-2}~.
\end{equation}
The first factor is the orbital phase precision that our method can improve to $\Delta\phi_{\rm int}\lesssim10^{-7}$. To be more thorough, we should also consider whether our method affects the other two factors. In the last factor, $T_{\rm tot}$ is the time difference between the first and the last data points, and it is parallel to our improvement here. The middle factor $n_{\rm data}$ is the inverse data density, namely how many orbital periods can one produce a precise measurement of the orbital phase. Our method may limit how small this number can be.

As described earlier, for slow pulsars our method requires extrapolating to the center of the interference pattern while only seeing a segment of it. When such segment is further away, a larger error is involved in this extrapolation. Let us be very conservative and demand to use only those pulses which cover the superior conjunction passage. Since the pulse duration is roughly $(1/100)$ of the pulse period, this only happens once every 100 periods. Thus our method requires $n_{\rm data}>100$. For an 8-hour orbit, this means around 10 data points per year. The upper-bound derived from the Hulse-Taylor binary has 1 data point per year, and can only be improved up to $26$ since each data point requires two weeks of observations \cite{HMY}. This is the only possible side-effect our method may have, and it does not make a big difference. Therefore, it is fair to just compare the orbital phase precisions. Our projected native precision is already 1 order of magnitude better than the actual precision from the Hulse-Taylor data after signal-to-noise analysis. It is likely that the upper-bound on $h_c$ can be pushed down by a few orders of magnitude.

In addition to this improvement, an interferometer is a generally powerful tool to probe the geometry of space-time. For example, if the companion has a large spin, there will be frame dragging effect. Without interference, this has been considered unmeasurable from timing information alone \cite{2006PhRvD..73f3003R}. We can estimate the sensitivity of our interferometer for frame dragging by considering a black hole with spin $a$ where $|a|=1$ means extremal. When the source is within the Einstein radius, its frame dragging effect can lead to an additional length difference between two strongly-lensed paths \cite{2006PhRvD..73f3003R},
\begin{equation}
\Delta l_{FD} \sim \frac{aM^2}{R_{\rm lens}}~.
\end{equation}
For an order one value of $a$ and the particular parameters we used in this paper, this is slightly less than one meter. In other words, it is about one wavelength at $GHz$. Although it is less than one part of a thousand of the total length difference in Eq.~(\ref{eq-pathdiff}), it leads to an asymmetry of the interference pattern in terms of time reflection about the superior conjunction. Since $\Delta\phi_{\rm amp}/\Delta\phi_{\rm int}\gtrsim1000$, the observable interference pattern is about thousands of peaks wide. Thus this effect is potentially observable. 

Finally, let us comment on the possibility to find a suitable binary for our method. The Square Kilometer Array projected the discovery of hundreds of compact pulsar binaries \cite{SKA}. Population synthesis suggests that in about $(1/5)$ of those, the companion shall be a black hole \cite{SchFer00}. If we are optimistic and say that an order one fraction of neutron stars have magnetospheres smaller than $10^6m$, then there will be hundreds of likely candidates. The most limiting requirement seems to be the inclination angle. We require the binary rotation axis to be almost orthogonal to the line of sight. For a Hulse-Taylor like orbit, such probability for that is roughly
\begin{equation}
\mathcal{P}\left(|\theta_{\rm tilt}|<\frac{R_{\rm lens}}{R_{\rm orbit}}\right) \sim \frac{2\pi\times (2R_{\rm lens}/R_{\rm orbit})}{4\pi} 
= 10^{-3}~.
\end{equation}
One might want to scan through different orbital sizes. For example, at $R_{\rm orbit}=10^7m$, the above probability improves to $10^{-2}$. However, at this distance or smaller, the pulse may not have a wide enough opening angle to cover the entire lens. Also, the remaining inspiral will not last longer than the pulsar lifetime, so we cannot expect to find any pulsar binary with a smaller orbit. Thus scanning over orbital sizes still gives about $10^{-3}$ probability in finding a suitable pulsar binary.

Given hundreds of binaries, a $10^{-3}$ chance cannot guarantee that we will find one, but we can still keep our hope high. On top of improving the bound on stochastic GW background, such a high precision measurement can also improve our knowledge about the pulsar and its companion in various ways \cite{JohGwi12,PenBro13,PenMac14}. \\ \ \\
{\bf Acknowledgments} 

We thank Jason Hessels, Lam Hui, Sean McWilliams and Roman Rafikov for discussions. This work is supported in part by the Foundation for Fundamental Research on Matter (FOM), which is part of the Netherlands Organization for Scientific Research (NWO).
\newcommand{\mnras}{MNRAS}   

\bibliography{gw}

\end{document}